\documentclass[10pt]{article}

\usepackage[utf8]{inputenc}
\usepackage[T1]{fontenc}

\usepackage{caption}
\usepackage{graphicx}
\usepackage{amsmath}
\usepackage{amssymb}
\usepackage{color}
\usepackage{simplemargins}

\settopmargin{2cm}
\setbottommargin{2cm}
\setleftmargin{2cm}
\setrightmargin{2cm}

\newcommand{\igj}[1]{\textcolor{black}{#1}}
\newcommand{\rk}[1]{\textcolor{black}{#1}}
\newcommand{\sj}[1]{\textcolor{black}{#1}}

\newcommand{\beginsupplement}{%
        \setcounter{table}{0}
        \renewcommand{\thetable}{S\arabic{table}}%
        \setcounter{figure}{0}
        \renewcommand{\thefigure}{S\arabic{figure}}%
        \captionsetup[figure]{labelfont={bf},name={Supplementary Figure}}
     }

\title{Intracellular Energy Variability Modulates Cellular Decision-Making Capacity \\ \vspace{0.2cm} \normalsize Ryan Kerr${}^1$, Sara Jabbari${}^1$, Iain G. Johnston${}^{2,3,*}$ \\ \vspace{0.2cm} \footnotesize ${}^1$ School of Mathematics \& Institute of Microbiology and Infection, University of Birmingham, United Kingdom \\ \footnotesize ${}^2$ Department of Mathematics, Faculty of Mathematics and Natural Sciences, University of Bergen, Norway \\ \footnotesize ${}^3$ Alan Turing Institute, London, United Kingdom \\ \footnotesize Correspondence to \url{iain.johnston@uib.no}}
\date{}

\begin{document}

\maketitle

\section*{Abstract}
Cells are able to generate phenotypic diversity both during development and in response to stressful and changing environments, aiding survival. The biologically and medically vital process of a cell assuming a functionally important fate from a range of phenotypic possibilities can be thought of as a cell decision. To make these decisions, a cell relies on energy dependent pathways of signalling and expression. However, energy availability is often overlooked as a modulator of cellular decision-making. As cells can vary dramatically in energy availability, this limits our knowledge of how this key biological axis affects cell behaviour. Here, we consider the energy dependence of a highly generalisable decision-making regulatory network, and show that energy variability changes the sets of decisions a cell can make and the ease with which they can be made. Increasing intracellular energy levels can increase the number of stable phenotypes it can generate, corresponding to increased decision-making capacity. For this decision-making architecture, a cell with intracellular energy below a threshold is limited to a singular phenotype, potentially forcing the adoption of a specific cell fate. We suggest that common energetic differences between cells may explain some of the observed variability in cellular decision-making, and demonstrate the importance of considering energy levels in several diverse biological decision-making phenomena.

k

%\noindent Please note: Abbreviations should be introduced at the first mention in the main text – no abbreviations lists. Suggested structure of main text (not enforced) is provided below.

%%-----------------------------------------------------------------------------------------------------------------
\section*{\label{sec:intro}Introduction}
%%-----------------------------------------------------------------------------------------------------------------

Biological cells are faced with many decisions during their existence. Genetically identical single cells in a population choose different phenotypic strategies for survival; genetically identical cells in developing multicellular organisms make decisions to follow different developmental pathways, and hence towards one of a diverse range of possible phenotypes. Across organisms, gene expression variability leads to cell-to-cell variations in mRNA and protein levels in genetically identical cells and can drive the generation of diverse phenotypes and strategies for survival \cite{cortijo2019widespread, kalmar2009regulated, ozbudak2002regulation, blake2006phenotypic, fraser2009chance}. The process of a cell assuming different functionally important fates from a range of phenotypic possibilities in response to or in anticipation of extracellular change, without genetic modifications, is known as a cellular decision \cite{bowsher2014environmental}.

In multicellular organisms, phenotypic heterogeneity has been observed in a diverse range of cell types \cite{balazsi2011cellular} for a wide range of cellular decisions, from seed germination \cite{johnston2018identification, mitchell2016variability, topham2017temperature} through the famous example of hematopoietic cell differentiation \cite{easwaran2014cancer, chang2008transcriptome}, to mosaic development of retinal cells in the \textit{Drosophila} eye \cite{wernet2006stochastic}. Waddington's famous `epigenetic landscape' \cite{waddington2014strategy} pictures these developmental decisions as bifurcating channels that a developmental `ball' can roll down to select different possible cell fate decisions; bifurcations in the landscape correspond to multistability, where a cell can support distinct, differentiated cell fates. These repeated differentation decisions allow, for example, human pluripotent stem cells to differentiate into all cell types in the human body \cite{park2008reprogramming}, while modern technology allows reprogrammed cells to move back `up' the epigenetic landscape \cite{park2008reprogramming, takahashi2007induction}. These cellular decisions are central to development and knowledge of their dynamics will offer useful applications in medicine and fundamental biology \cite{mahla2016stem}.

Single-celled organisms also embrace the advantages of diverse cell behaviours. Cells in their natural environment have to deal with the challenges presented by changes in extracellular conditions. These may include temperature changes, pH variability, nutrient limitation or, in some cases, the presence of antibiotics. To overcome such environmental fluctuations, genetically homogenous cells can generate phenotypic diversity in order to increase the probability that some members of the population will survive \cite{hughes2007experimental, levy2016cellular, mitchell2009adaptive}. This variability may enable a cell to support multiple, distinct phenotypes \cite{acar2008stochastic, suel2006excitable, wernet2006stochastic} potentially helping a cell handle different environmental conditions; the resulting population heterogeneity may even increase the overall fitness of the species \cite{acar2008stochastic, smits2006phenotypic}. 

In microbiology, diversity and cell decision-making underlie a range of biologically and medically important behaviours. Large-scale studies have revealed the generation of phenotypic diversity in yeast through noisy gene expression, potentially benefitting the population in varying and often stressful environments \cite{blake2006phenotypic, newman2006single}. The decision for bacteria to sporulate into robust endospores is a survival mechanism used by diverse genera including \textit{Bacillus} and \textit{Clostridium} \cite{phillips2002bacillus}, contributing to foodborne disease and food spoilage \cite{andersson1995problems, wells2016bacterial}. The decision to become a persister cell (a phenotype more tolerant to external stress, including antibiotics), is made in several bacterial species \cite{balaban2004bacterial, cohen2013microbial, conlon2016persister, kussell2005bacterial} and can have a dramatic impact on the efficacy of treatments \cite{fauvart2011role, lewis2010persister, mulcahy2010emergence, zhang2012targeting}.

Many of the mechanisms behind cellular decision-making in eukaryotes and prokaryotes remain poorly understood. Regulatory networks, representing the interactions between genes that govern these decisions, are often used to summarise our knowledge \cite{karlebach2008modelling}. Typically, edges in a schematic network illustrate processes such as transcription and translation and nodes represent genes. However, this coarse-graining can often omit a substantial amount of important detail. In particular, the fact that the processes represented by these edges are energy dependent (Figure \ref{fig:gene12}) is rarely considered. Transcription and translation require a substantial ATP budget \cite{das2010connecting, kafri2016cost}, so there exists a core energy dependence in the dynamics of gene regulatory networks, potentially affecting the decisions supported by a given cell.

This energy dependence is important because different cells, particularly in microbiology, can have substantially different levels of available energy. Energy variability has been observed within genetically-identical \textit{Escherichia coli} bacteria cells in a population, where absolute concentrations of intracellular ATP were spread over at least half an order of magnitude, in a skewed distribution around $1.54 \pm 1.22$mM \cite{yaginuma2014diversity}. Substantial intracellular ATP variability has also been observed in other branches of life, including yeast \cite{takaine2019reliable}, HeLa cells \cite{yoshida2016bteam} and plants \cite{de2017atp}. Clearly, the dynamics, and thus potentially the outputs, of cellular decision-making networks may vary between cells due to these diverse energy levels.

Previous work has shown that energy availability can modulate the stability of decision-making circuits in eukaryotic cells \cite{johnston2012mitochondrial}, supported by experimental observations linking, for example, mitochondrial content and quality (supporting high, stable ATP supply) to less differentiated cell outcomes \cite{katajisto2015asymmetric, schieke2008mitochondrial}. However, the broader influence of energy variability on the behaviour of these circuits remains to be revealed. Here, we use a general regulatory network to model a genetic architecture observed as a decision-making motif across organisms. By including the effects of energy variability on physiological processes, we explore how the behaviour of the system will change when there is a divergence in intracellular energy and what effects it has on the ability of cells to make diverse decisions.

%%-----------------------------------------------------------------------------------------------------------------
\section*{\label{sec:model}Methods}
%%-----------------------------------------------------------------------------------------------------------------
\subsection*{\label{sec:model_orig}Regulatory Architecture}
\begin{figure}%[h]
	\centering
	\includegraphics[width=16cm]{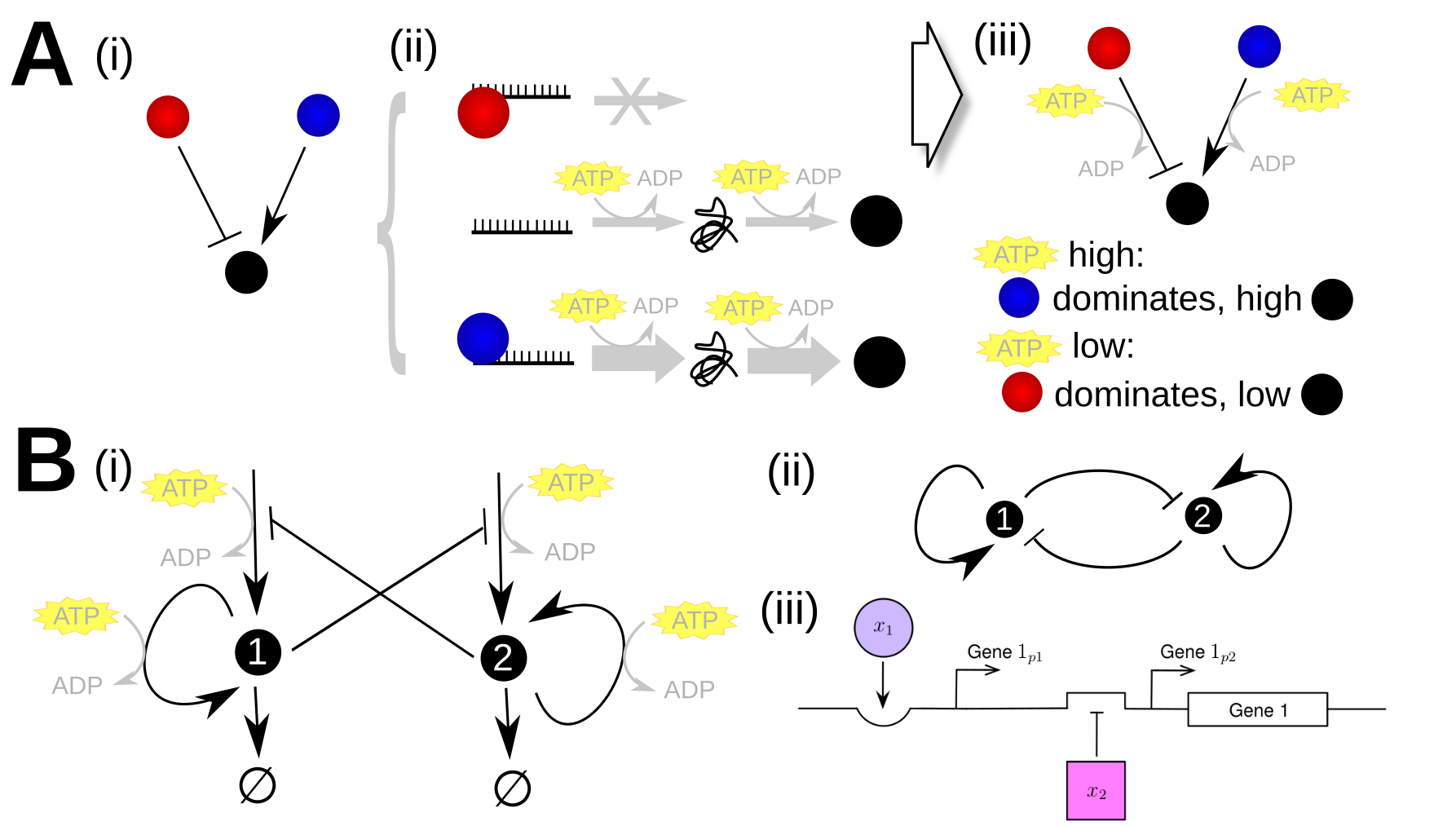}
	\caption{\textbf{Energy dependence in model genetic architectures.} \textbf{A.} Energy dependence in gene regulation. In an example schematic regulatory network (i), genes have positive and negative regulatory interactions. The expression of these genes, and thus their regulatory interactions, rely on transcription and translation, processes with substantial ATP requirements (ii). The dynamics, behaviour, and states supported by a regulatory network are thus expected to be ATP-dependent (iii). In this example, when ATP levels are high, activation by a positive promoter (blue) may lead to high levels of expression of a downstream target (black). However, when ATP levels are low, there may be insufficient energy to support this increased expression level, and the system's behaviour is instead dominated by a repressive interaction (red). \textbf{B.} The regulatory network for the general system considered in our model. The system involves two genes which self-activate and cross-repress, with both sets of processes being ATP-dependent (i). The energy dependence is often neglected in shorthand network representations of gene regulatory networks (ii). (iii) shows a possible arrangement of regulatory regions for one gene in the system that accomplishes this toggle-switch control.}
	\label{fig:gene12}
\end{figure}

We consider a well-known cellular decision-making architecture consisting of two genes, their respective protein products, and the regulation occurring through their interactions \cite{huang2007bifurcation, johnston2012mitochondrial}. To study a general model of cellular decision-making we designate the genes `Gene 1'\hspace{0.01cm} and `Gene 2'\hspace{0.01cm}, as displayed in Fig. \ref{fig:gene12}B. Protein 1 and Protein 2 (expression levels $x_1$ and $x_2$) are expressed by Gene 1 and Gene 2 respectively, which then apply feedback control to the system through self-activation and cross-repression; both proteins also degrade. We define a cell fate as the level of these proteins at steady state. 

We construct the governing ordinary differential equations (ODEs) with guidance from the literature \cite{huang2007bifurcation}. We coarse-grain gene expression dynamics, avoiding an explicit representation of mRNA levels, and focussing on the levels of the protein products. The expression of each gene has contributions from self-activation (via the gene's product activating a conditional promoter) and from a constitutive promoter (which can be repressed by the other gene's product), Fig. \ref{fig:gene12}B. For Gene $i$, the maximum expression level of the conditional promoter for Gene $i$ is $a_i$, and the basal expression level of the constitutive promoter is $b_i$. Each protein product is degraded with rate $k_i$. Overall we then obtain Equations \eqref{ODE1}-\eqref{ODE2} with initial conditions \eqref{ODE-ics}, previously studied by Huang et al. \cite{huang2007bifurcation}, with $x_{1,0}, x_{2,0} \in \mathbb{R}$ being some initial levels of proteins $x_1$ and $x_2$.

\begin{equation}\label{ODE1}
	\frac{\mathrm d x_1}{\mathrm dt} = a_1 \frac{x_{1}^{n}}{\theta_{a_{1}}^{n}+x_{1}^{n}}+b_{1} \frac{\theta_{b_{1}}^{n}}{\theta_{b_{1}}^{n}+x_{2}^{n}} - k_{1}x_{1},
\end{equation}
\begin{equation}\label{ODE2}
	\frac{\mathrm d x_2}{\mathrm dt} = a_2 \frac{x_{2}^{n}}{\theta_{a_{2}}^{n}+x_{2}^{n}}+b_{2} \frac{\theta_{b_{2}}^{n}}{\theta_{b_{2}}^{n}+x_{1}^{n}} - k_{2}x_{2},
\end{equation}
\begin{equation}\label{ODE-ics}
	x_{1}(0) = x_{1,0}, \ x_{2}(0) = x_{2,0}.
\end{equation}

The additive terms in Equations \eqref{ODE1}-\eqref{ODE2} reflect, from left to right, self-activation up to a maximum level of $a_i$, cross-repression down from a basal level of $b_i$, and degradation. The interaction processes are represented by Hill functions, with $n$ and $\theta$ parameters determining the steepness and inflection point of the sigmoidal curves, respectively; $\theta_{a_{i}}$, $\theta_{b_{i}}$ can be interpreted as the dissociation constants of the activator and inhibitor regulatory proteins to the promoter regions, respectively.

Huang \emph{et al.} suggest a default example parameter set \cite{huang2007bifurcation}, with $a_i = b_i = k_i = 1, \theta_{a_i } = \theta_{b_i } = 0.5, n = 4$ ($i = 1, 2$). These parameters give rise to a tristable system, where, depending on initial conditions, the steady state can take one of three values. Interpreting these protein level values as cell fates, this tristability corresponds to three possible cell fates for an organism and therefore, for example, a wider decision-making landscape compared to a monostable system.

\subsection*{\label{sec:model_energy}Energy Dependence}
Each step in transcription and translation requires ATP (Figure \ref{fig:gene12}), so we enforce that the rates of the corresponding gene expression processes in our model are dependent on an ATP concentration parameter. To this end, we first transform the parameters $a_i$ and $b_i$ to $a_{i} \rightarrow \lambda a_{i}$ and $b_{i} \rightarrow \lambda b_{i}$ ($i = 1, 2$) with $\lambda$ being some function of intracellular energy budget. For simplicity and generality, we ignore the possible energy dependence of degradation as being of lower magnitude than these constructive processes. We will consider $\lambda$ values between 0 and 1. 

$\lambda$ is considered a function of scaled intracellular energy budget $A^{*}$; a ratio of a cell's free energy availability to some maximal possible value. The choice of function to model $\lambda$ depends on the biological process being considered: for example, free energy available from ATP depends on ATP:ADP ratio and other metabolic factors. We follow das Neves \emph{et al.} \cite{das2010connecting}, who found a sigmoidal relationship between the total transcription rate in a cell and ATP concentration. As we are primarily concerned with the ATP dependence of gene expression, we modelled $\lambda$ with a sigmoidal curve, yielding monotonically increasing rates as energy increases (adoption of a linear rather than a sigmoidal relationship between scaled energy and dynamic rates did not change our qualitative observations, Supplementary Fig. S1). Specifically, we use Equation \eqref{lambda_sig_eqn}, where  $s_1=16, s_2=-8$ are chosen to produce the monotonically increasing curve for $\lambda(A^{*})$ in Supplementary Fig. S2.

\begin{equation}\label{lambda_sig_eqn}
    \lambda(A^{*}) = \frac{1}{1+e^{-(s_1A^{*}+s_2)}}.
\end{equation}

For biological reference, Supplementary Fig. S2 displays the known range of intracellular ATP concentration of \textit{E. coli} ($1.54 \pm 1.22$~mM \cite{yaginuma2014diversity}). At the lower bound of ATP concentration, $320~\mu$M, $\lambda$ is small, equating to a low maximum expression level. For the upper bound, $2760~\mu$M, $\lambda$ is at its maximum value and represents maximal expression level. Concentrations below $320~\mu$M are outside the detected \textit{E. coli} ATP range \cite{yaginuma2014diversity}, so we take this to be the critical intracellular energy to continue as a living cell and therefore may represent non-living or dying cells. Overall we thus obtain:
    
\begin{equation}\label{ODE1_ATP}
	\frac{\mathrm d x_1}{\mathrm dt} =  \lambda(A^{*})a_1 \frac{x_{1}^{n}}{\theta_{a_{1}}^{n}+x_{1}^{n}}+ \lambda(A^{*})b_{1} \frac{\theta_{b_{1}}^{n}}{\theta_{b_{1}}^{n}+x_{2}^{n}} - k_{1}x_{1},
\end{equation}

\begin{equation}\label{ODE2_ATP}
	\frac{\mathrm d x_2}{\mathrm dt} =  \lambda(A^{*})a_2 \frac{x_{2}^{n}}{\theta_{a_{2}}^{n}+x_{2}^{n}}+ \lambda(A^{*})b_{2} \frac{\theta_{b_{2}}^{n}}{\theta_{b_{2}}^{n}+x_{1}^{n}} - k_{2}x_{2}.
\end{equation}

Time-dependent and steady state numerical solutions of Equations \eqref{ODE1_ATP}-\eqref{ODE2_ATP}, with Equation \eqref{lambda_sig_eqn}, were generated in Matlab (R2018a; Mathworks) using the ode45 and fsolve solvers; ode45 is a non-stiff first-order ordinary differential equation solver which implements a numerical method of the Runge-Kutta family and fsolve is a non-linear solver for non-linear equations which uses a Trust-Region algorithm, with each iteration using the method of preconditioned conjugate gradients. Prescribed initial conditions covered a grid of $x_1$ and $x_2$ at equally spaced points when calculating numerical solutions; bifurcation diagrams and heatmaps were constructed using the numerical solutions. In the case where multiple degenerate solutions are found, attractor quantity was confirmed using both Maple (2017) and Matlab (R2018a; Mathworks). All scripts used for this study are openly accessible through \url{https://github.com/kerr-mathbio/energy_variability_decision_making.git}.

For each parameter set, the stable steady states of the system correspond to cell fates. We do not consider unstable or metastable steady states to represent cell fates as small perturbations in expression due to noise will move the system towards a stable point; we consider both unstable and metastable steady states as `functionally unstable' in our study. If multiple stable states exist, the level of proteins $x_1$ and $x_2$ at each steady state determine the different phenotypes a cell can generate, as shown by Huang \emph{et al.} \cite{huang2007bifurcation}; in their study, high $x_1$ and low $x_2$ corresponded to a progenitor cell differentiating into an erythroid cell, rather than a myeloid cell. 

As we are modelling the observed interactions in Fig. \ref{fig:gene12}\rk{B} rather than biochemical interactions, the analysis of our results considers the qualitative behaviour rather than observing quantitative results. From this point, we set $a = a_1=a_2$, $b = b_1=b_2$, $k = k_1=k_2$, $\theta_a = \theta_{a_1} = \theta_{a_2}$, $\theta_b = \theta_{b_1} = \theta_{b_2}$ and each section will explicitly state the fixed and varied parameters for the work that follows.

\section*{Results}

Our model system generally exhibits dynamic behaviour that, starting from some initial condition $(x_{1,0}, x_{2,0})$, converges to a particular steady state characterised by values $(x_1, x_2)$. A limited number of these steady states exist for a given parameterisation of the system; these are \emph{attractors}, so called because initial conditions are `attracted' towards these stable states. A range of initial conditions will converge to the same attractor; this range of $(x_{1,0}, x_{2,0})$ values is the \emph{attractor basin} of that attractor.

We view distinct attractors, with distinct patterns of protein levels, as distinct cell fates. The attractor basin of each corresponds to the range of cell states that will return to that cell fate. This range makes each fate robust to fluctuations: for example, a small change in the level of one protein in one attractor state will likely still fall within that attractor's basin, so the system will return there.

In the following sections, we consider the number and properties of distinct cell fates (attractors) supported by the system at different energy levels. As we increase energy levels we often observe \emph{bifurcation}, where one stable state transitions to two new states (increasing the number of options for cellular decision making). Here, we do not consider the dynamics of switching between attractors (moving from one cell fate to another); in biology this can be accomplished through noisy or controlled external influences \cite{paulsson2004summing, pedraza2005noise, thattai2001intrinsic}.
  
\subsection*{Intracellular Energy Budget Modulates Cellular Decision-Making Landscapes}
We first sought to understand how increasing energy availability changes the number of decision-making options available to the cell. To this end, we explored the bifurcation behaviour of the steady-state protein levels in our model as we changed energy availability. As our parameterisation for now imposes a symmetric structure on the phase space of steady-state protein levels, we begin by simply using the diversity of state level $x_2$ levels to illustrate the range of attractors present in the system for given energy levels (example in Supplementary Fig. S3). Throughout this section, $\theta_a$, $\theta_b$ and $n$ are fixed at their default values, and parameters $a$, $b$ and $k$ are varied. 

\begin{figure}%[h]
	\includegraphics[width=\textwidth]{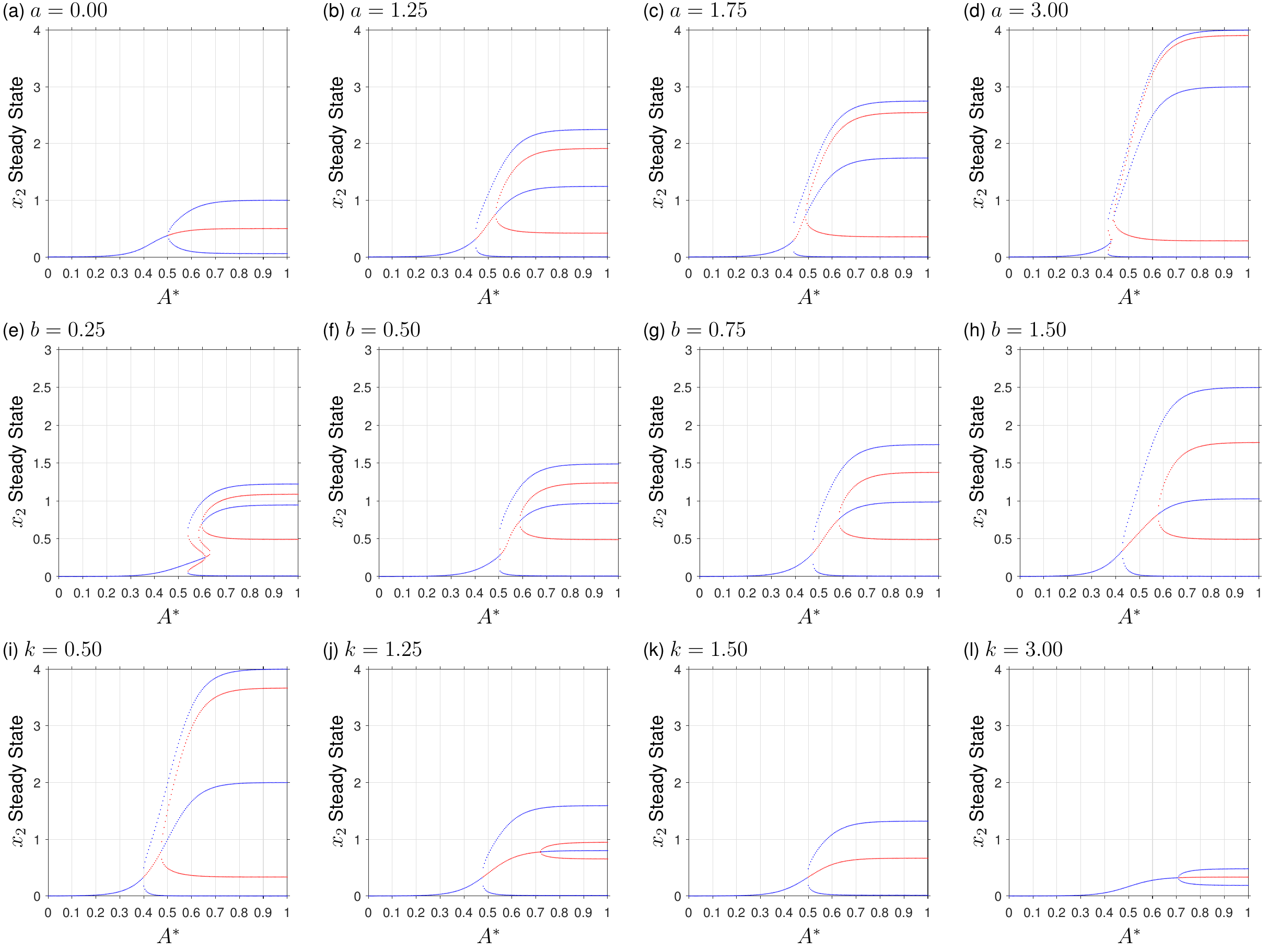}
	\caption{\textbf{Attractor basins with increasing energy availability $A^*$.} Bifurcation diagrams show increasing values for: (a)-(d) maximum expression level from conditional promoter; (e)-(h) constitutive promoter level; (i)-(l) degradation. Plots display $x_2$ values for stable steady states (blue circles) and unstable steady states (red circles) over $A^{*} \in [0, 1]$ in steps of size $5\times 10^{-3}$.}
	\label{fig:bif_diag}
\end{figure}

We explored the emergence of different attractors (hence, the emergence of different cell fates that can be decided between) with energy level $A^*$ under a variety of different parameterisations. A clear general trend emerged, whereby the number of distinct attractors supported by the system increases with increasing energy availability (Fig. \ref{fig:bif_diag}). There is thus an increased diversity of stable protein states, and hence an increased number of options for cellular decisions, if energy levels are higher. The separation of these attractors also increased with increasing energy, with the attractor basin associated with the intermediate attractor becoming wider, in agreement with \cite{johnston2012mitochondrial,huang2007bifurcation}. This suggests a biological stabilisation of states characterised by intermediate protein levels (see Fig. \ref{fig:bif_diag}).

The quantitative properties of these attractor states vary with model parameters. Increased promoter activity $a$ and $b$ has the effect of increasing the separation of protein levels in distinct states, likely due to a simple elevation of maximum level. Correspondingly, increased degradation activity $k$ has the opposite effect, diminishing the differences between distinct states. In particular, high values of $k$ prevent the emergence of tristability, limiting the system to bistability even at high energy levels (Fig. \ref{fig:bif_diag}(k-l)).
 
\begin{figure}%[htb]
	\centering
	\includegraphics[width=\textwidth]{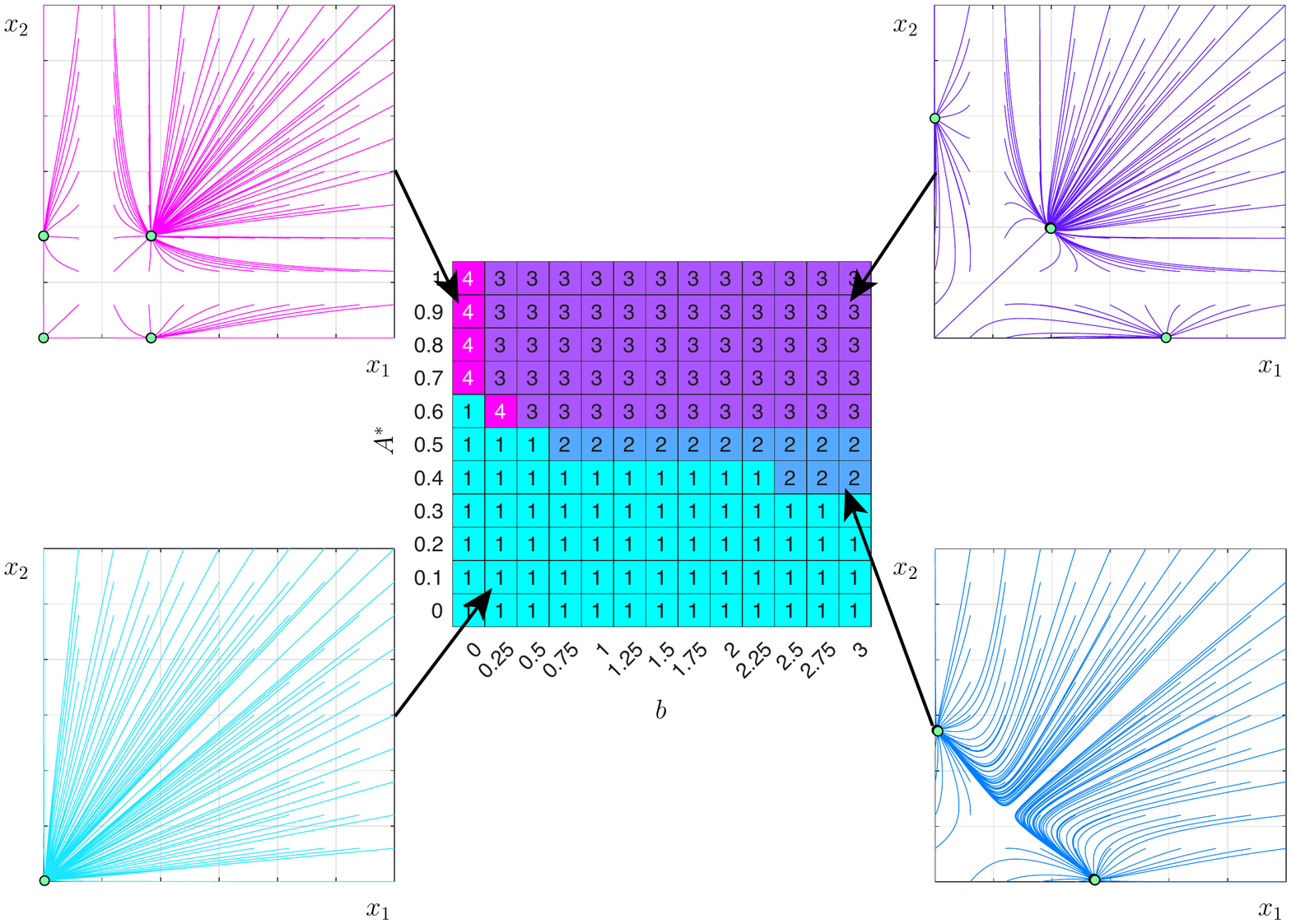}
	\caption{\textbf{Decision-making capacity depends on cellular energy and dynamic parameters.} Heatmap for $a=1$, $b \in [0, 3]$ and $A^{*} \in [0, 1]$, with all remaining parameters fixed at their default values. Inset figures present examples of the attractor landscape with solution trajectories (coloured) and stable steady states (green circles with black circumference). The $4$ categories are, when viewed in portrait orientation, $1$ stable steady state (bottom left, turquoise), $2$ stable steady states (bottom right, blue), $3$ stable steady states (top right, purple) and $4$ stable steady states (top left, pink).}
	\label{fig:heatmap_examples}
\end{figure}

\begin{figure}%[htbp]
  \includegraphics[width=\textwidth]{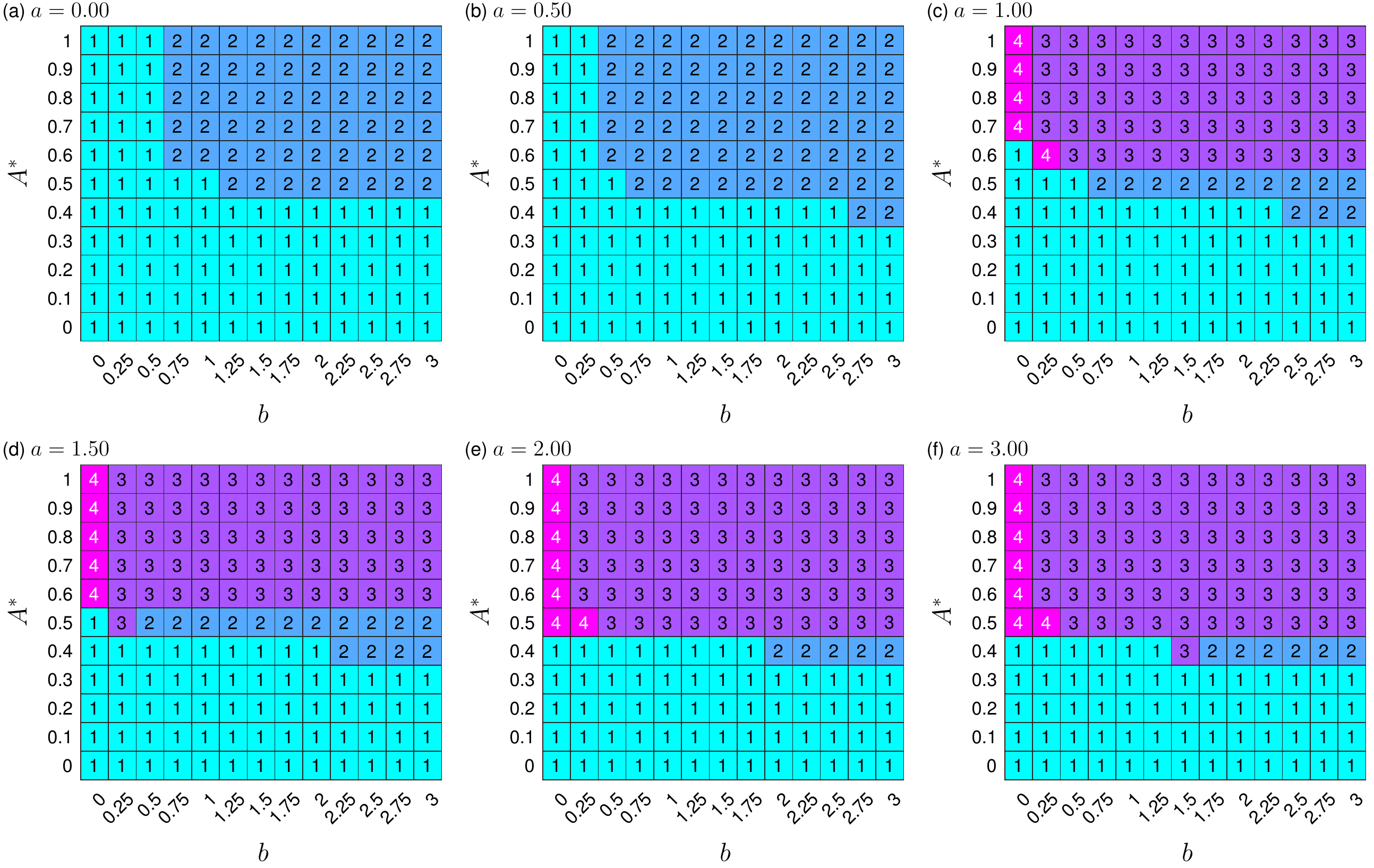}
	\caption{\textbf{Decision-making landscapes depend on energy availability and dynamic rate parameters.} Panels (a)-(f) display heatmaps for $6$ increasing values of $a$, when $n=4$. Each panel exhibits the number of stable steady states for combinations of $b \in [0, 3]$ and $A^{*} \in [0, 1]$. Throughout (a)-(f), all parameters, except $a, b$ and $A^{*}$, are fixed at their default values.}
	\label{fig:heatmaps}
\end{figure}

The bifurcation dynamics display some subtle variation in different cases. For example, at high levels of $a$ (the maximum expression level of the conditional promoter), increasing energy availability drives the system through re-entrant behaviour, where the number of attractors (decision options) runs from 1 to 3 then down to 2 before returning to 3 (see Supplementary Figs. S4 and S5 for detailed views of this behaviour). We shall see below (for example in Figs. \ref{fig:heatmap_examples} and \ref{fig:heatmaps}) that similar behaviour is also observed in other regions of parameter space. 

Generally, a single stable branch exists below an $A^{*}$ threshold. Biologically this implies that a single cell fate exists below an intracellular energy level, restricting the decision-making ability of a cell under this architecture. Once intracellular energy budget exceeds a threshold, multiple steady states are present, enabling cellular decision-making. All figures demonstrate that, for our decision-making architecture, an increase in $A^{*}$ generally has the effect of accumulating stable branches in the bifurcation diagrams. This means a broader decision-making landscape for a cell when it has high intracellular energy budget. If degradation, $k$, is too high the central stable attractor does not exist and the system is limited to bistability. Therefore, the decision-making capabilities of a cell would be reduced if the degradation level is sufficiently high. Re-entrant stable steady state behaviour exists for certain parameter sets, suggesting a potential optimum intracellular energy range for phenotypic diversity. This may give a cell an optimal range of intracellular energy levels to support an increased variety in cell phenotypes, enabling superior adaption to stochastic extracellular environmental changes. This behaviour could also have negative consequences for a cell due to phenotypes becoming unsupported with slight variances in energy, forcing a cell to establish a different phenotype which, due to being a time dependent process, may not be performed quick enough to tolerate an environmental change.

In Fig. \ref{fig:heatmap_examples} we show this `phase portrait' of attractor landscapes for varying energy $A^*$ and constitutive expression $b$, along with example structures of the corresponding attractor basins. The general observation of increasing number of fate options with energy holds throughout. There is a separatrix line in this phase portrait between regions supporting a single attractor basin and those supporting more than one. For a given dynamic parameter, this separatrix is crossed as energy availability increases, then further increase has the effect of expanding the attractor basin associated with the state where both genes are expressed symmetrically (as in \sj{Johnston \textit{et al.}}\cite{johnston2012mitochondrial}).

%\subsection{Energy Modulates Cellular Decision-Making Landscapes}
%\label{sec:landscape}

%We next sought to expand these specific results to a broader and more general picture of the relationship between dynamic parameters, energy availability, and decision-making potential. To this end, we explored how the number and nature of distinct attractor basins depend jointly on energy $A^*$ and other governing parameters.

What is the physical intuition behind these structural changes with energy and dynamic parameters? Start with the low activity, low energy case. Here there is simply not enough cellular `production capacity' -- the ability to express genes -- to allow dynamic signalling interactions between gene products. The system is limited to low levels of expression for both products, favouring a single, low-expression state.

As we now increase $b$, the basal expression rate increases, and the high coefficient associated with the cross-repression term induces very high antagonism -- any balance between the gene products is highly unstable. The system invests all available energy getting to either one state or another. The low energy levels and correspondingly low expression levels mean that `small number' effects are important: any small perturbations from symmetric expression will be amplified to favour one gene over the other.

In the alternative case, where we have low basal expression but high energy, we have substantial `production capacity' that can now support an energetically more demanding intermediate state. Higher expression levels mean that fluctuations can be buffered and intermediate states are more stable. Here, there is constant tension between genes, requiring high energy levels; the fixed states are also still supported. In order to favour either gene product we need enough initial protein to jump-start the system (otherwise we reach the trivial zero or near-zero attractor).

This latter case remains true in the final case of high energy and high basal expression, except here the zero or near-zero attractor is unstable because from here, the substantial `production capacity' means that some protein can always be produced, which is likely to drive the system towards one of the non-trivial attractors.

We next explored this behaviour varying both promoter levels $a$ and $b$ together (Fig. \ref{fig:heatmaps}). At low values of the conditional promoter activity ($a \leq 0.5$) the system is restricted to mono- or bistability for all values of $b$ and a single stable attractor exists below an $A^{*}$ threshold. High values of $a$ support the greater diversity of stable states previously seen in Fig. \ref{fig:heatmap_examples}. For high $a$ and some values of $b$, we again see the re-entrant behaviour from the previous section, with the 4-attractor state supported only at some intermediate values of energy availability. Illustrations of the nature of these re-entrant steady states are given in Supplementary Figs. S6 and S7.

\igj{This re-entrant behaviour (Supplementary Figs. S4-S7) merits further analysis. The reduction from 4 to 3 stable states with increasing energy is likely due to the destabilisation of the near-zero attractor (Supplementary Fig. S6), where expression levels of both genes is low. Above an energetic threshold, a baseline of production will be constantly occurring, and if the activity $a$ of the conditional promoter is high enough, any small increases in protein level can rapidly self-amplify and drive the system to an attractor with higher expression of protein products. The 4-attractor system will then only be supported at low $b$ and sufficiently low energies, as we observe. When the number of attractors (decision options) runs from 1 to 3 then down to 2 before returning to 3, the central stable attractor is destabilised as energy increases (Supplementary Fig. S7). This is likely due to a higher contribution from the constitutively expressed promoter as a proportion of the overall expression when energy increases, enhancing the antagonism, increasing the overall expression of both proteins and therefore driving the system to one state or the other as the central state cannot be sustained.}

In summary, when $a$ is small, the system is limited to mono- or bistability, which may equate to a limited diversity in phenotypic possibilities for a cell. Increasing $a$ lowers the separatrix and decreases the $A^{*}$ threshold for tristability. Biologically, this suggests that for our decision-making architecture, as the maximum expression level of the promoter under activator control increases, less energy is required for an expansion in a cell's decision-making landscape. Independent of the value of $a$ and $b$, a singular stable steady state always exists at low energy, in theory limiting a cell to generating a singular phenotype.

\subsection*{Cooperativity Effects on System Behaviour}
We next asked how these relationships between attractor basin structure (number of decision options) and energy depends on the cooperative nature of our model interactions (represented by the Hill coefficient $n$ in our governing equations). We found that decreasing the Hill coefficient to $n = 3$ (Supplementary Fig. S8) led to some minor rearrangements of the phase portrait, decreasing to $n=2$ (Supplementary Fig. S9) does not display any re-entrant behaviour, while the quasi-linear case of $n=1$ (Supplementary Fig. S10) removed any diversity in attractor basin structure for the parameters we consider.

Therefore, if there is noncooperative binding ($n = 1$) of the proteins, the number of stable states is limited to one. This would relate to no decision-making capability for any cell, independent of intracellular energy level or gene expression levels, in the regimes we are considering. In contrast, if cooperativity is reduced from $n=4$ (default value) to $n=2$ or $n = 3$, the observed behaviour is qualitatively similar, suggesting that a limited degree of nonlinearity in cellular interactions is sufficient to support decision-making and that energy dependence remains important even in these limited cases.

\subsection*{Regulatory Protein Binding Strength Can Limit System Behaviour}
To understand how the binding strength of regulatory proteins affects the decision-making abilities of a cell, the influence of $\theta_a$ and $\theta_b$ on attractor basin structure was also analysed (Supplementary Fig. S11); $\theta_a$ and $\theta_b$ can be interpreted as the dissociation constants of the activator and inhibitor regulatory proteins to the promoter regions, respectively. Increasing $\theta_a$ ($\theta_b$) corresponds to decreasing the binding affinity of the corresponding protein to the relevant promoter site, thus requiring more protein to achieve the equivalent levels of activation (inhibition). 
%Increasing $\theta$ stretches each Hill function. Therefore, increasing $\theta_a$ decreases the affinity of regulatory interaction - $\theta$ can be observed in Michaelis-Menton kinetics as $text{K}_{d}=1/K_a$ where K_{a} = k_bind/k_unbind is the association constant. If K$_{d}$ decreases --> K_a increases --> the affinity for the binding increases --> smaller the K_d value, the greater the binding affinity of the ligand for its target. 

Changing $\theta_a$ changes the $A^*$ threshold required for tristability; there is a `sweet spot' in $\theta_a$ that supports tristability at higher $A^*$ values. This behaviour is not observed for $\theta_b$; for low $\theta_b$, single, double, and triple stable states are observed at different energy levels, while for high $\theta_b$ the system is limited to a single attractor.

Hence, the relationship between energy availability and decision-making capacity of the system depends differently on the interactions encoded by $\theta_a$ and $\theta_b$. There is an optimal range of $\theta_a$ for tristability, and outside of this range it is more probable for the system to be bistable, but it can also be limited to monostability. This suggests that a stronger activator binding strength is required for a cell to be less sensitive to changes in the number of activator proteins modulating expression levels. The optimal range for tristability is increased when parameter $\theta_b$ is varied. A cell would be less sensitive to changes in the number of inhibitor proteins modulating expression for both weak and strong binding strengths of the inhibitor protein. When the inhibitor binding strength becomes too weak the system is limited to a single stable state, removing the decision-making ability for a cell.

\section*{Discussion}
%% summary of paper findings
We have used a simple and general regulatory model, representative of the core of several known decision-making motifs \cite{feng2012new, huang2009non, huang2007bifurcation, okawa2016generalized, zhou2011understanding}, to explore how the decision-making capacity of a cell depends on the energy available to fuel the processes involved in this decision making. \igj{Our theory predicts that, across a broad range of biological contexts, differences in cellular energy levels will cause differences in the dynamics and stable outcomes of cellular decision-making, via modulation of the expression rates of interacting regulatory genes. More precisely, increased ATP levels both support the ability to transition into multiple stable states, and increase the separation between the attractors as energy increases, stabilising the decisions that can be made (Fig. \ref{fig:bif_diag}(a)). Concurrently, without a sufficient energy supply, decision-making circuits lose the capacity to select different phenotypes (Figs. \ref{fig:bif_diag}(a) and \ref{fig:heatmaps}(a)). Higher energy levels allow, and stabilise, separate and intermediate states. }

\igj{This observation that different levels of ATP may induce different cellular behaviours comes as increasing cell heterogeneity in ATP levels is being experimentally characterised.} \rk{Intracellular ATP has been quantified to vary between 0.32-2.76mM in a bacterial population \cite{yaginuma2014diversity}, 3.7-3.9mM in HeLa cells \cite{yoshida2016bteam}, 3-4mM in living yeast cells \cite{takaine2019reliable} and over a several-fold change occurs in different tissues of a plant \cite{de2017atp}. The connection between energy variability and cellular decision-making has also been discussed previously \cite{conlon2016persister, shan2017atp}, further supporting the hypothesis that energy diversity may modulate the number of regulatory states supported by a cell, playing an important causal role in inducing variability in cellular decision-making.}

%% applications to other decision-making circuits - natural
\rk{The architecture we consider (Fig. \ref{fig:gene12}\rk{B}) can describe the behaviour of many naturally-occurring decision-making circuits, and our findings that energy availability influence this behaviour thus provides testable predictions for this class of systems. \igj{Accordingly, our prediction that phenotypic decisions depend on ATP} is already supported by several experimental observations, including those observing mitochondrial influence on stem cell fate \cite{katajisto2015asymmetric, khacho2018mitochondria, schieke2008mitochondrial} and phenotypic diversity \cite{guantes2015global}. \sj{\igj{Our hypothesis can further be readily tested by, for example, tracking prokaryotic phenotype or eukaryotic fate decisions in a culture exposed to the same stimulus but titrated with different external concentrations of ATP.}} }

Setting $a_1 = a_2 = 0$ reduces the system to consist of a bistable switch with cross repression, known as a toggle-switch. This motif is found at the core of several known decision-making architectures throughout biology, with an extensive body of associated literature \cite{bokes2013transcriptional, cherry2000make, lipshtat2006genetic, loinger2007stochastic, perez2016intrinsic, tyson2003sniffers, verd2014classification, warren2005chemical}. \rk{This motif famously occurs in the $\lambda$-bacteriophage switch from lysogenic to lytic state in \emph{Escherichia coli}. The lysis-lysogen decision can generally be described as being determined by two genes, \emph{cI} and \emph{cro}, which mutually cross-repress through their products; the outcome of high \emph{cI} (\emph{cro}) expression is \emph{cro} (\emph{cI}) repression and \emph{cI} (\emph{cro}) transcription, resulting in the lysogenic (lysis) state \cite{ackers1982quantitative, atsumi2006synthetic, aurell2002epigenetics, ptashne2004genetic}. The switch from lysogenic to lytic state occurs when the \emph{cI} product is cleaved by RecA, a result of the response to DNA damage, derepressing \emph{cro}, increasing Cro production and turning off \emph{cI} \cite{ackers1982quantitative}. The cross-repressing motif is observed in eukaryotic development \cite{huang2009non, zhou2011understanding}, where, for example, cross-repressing mammalian transcription factors Cdx2 and Oct4 govern the branch point between the trophectoderm and the inner cell mass in pluripotent embryonic cells \cite{niwa2005interaction}, and the bistable p42 MAPK/Cdc2 system controls maturation of the \textit{Xenopus} oocyte \cite{xiong2003positive}.} \rk{The situation $a \not= 0$, including self-promotion in the regulatory motif, describes more architectures such as B cells promoting an antibody class switch \cite{muto2010bach2}, gene regulatory networks with slow promoter kinetics \cite{al2019multi} and cell-fate development and differentiation in eukaryotic cells \cite{andrecut2011general, becskei2001positive, feng2012new, folguera2019multiscale, okawa2016generalized, wu2017engineering}. A famous example is the GATA-1-PU.1 system controlling hematopoietic stem cell differentiation \cite{al2019multi, bokes2009bistable, duff2012mathematical, huang2007bifurcation, johnston2012mitochondrial, roeder2006towards, tian2014mathematical}.} \rk{We predict that ATP variability will influence fate decisions across this broad variety of systems, organisms and branches of life. Higher ATP levels will promote undifferentiated states, and the propensity to choose different cell fates will be intrinsically affected by ATP supply (in addition to extrinsic signals that may also be linked to nutrient and energy availability).}

A particularly pertinent example of a cellular decision that may exhibit ATP dependence is the formation of bacterial persister phenotypes in the presence of antibiotics. One regulatory motif involved here is the \textit{hipBA} module, which expresses HipA (toxin) and HipB (antitoxin), which can form a complex that represses their own expression. Rotem \emph{et al.} \cite{rotem2010regulation} found that cells become dormant if the level of HipA exceeds a threshold, and as the level of HipA exceeded the threshold further, it determined the duration of dormancy. This is similar to our model and architecture where the protein levels, $x_1$ and $x_2$, feed back into the system and establish the generated cell fate; we would then predict that intracellular energy budget is a factor involved in the persister decision. Further connections between intracellular energy budget and the persister cell decision have been observed in both \textit{S. aureus} \cite{conlon2016persister} and \textit{E. coli} \cite{shan2017atp}.

%% applications to other decision-making circuits - synthetic
%\rk{Our architecture also has a well-known presence in synthetic biology, having been engineered and artificially implemented in \textit{Escherichia coli} \cite{gardner2000construction} and in mammalian cells \cite{kramer2004engineered}. In \emph{E. coli}, this artificial circuitry consists of genes \textit{lacI} and \textit{$\lambda$cI} that cross-repress, enabling cells to occupy one of two states \cite{gardner2000construction}. This was extended to be interfaced with the SOS signalling pathway \cite{kobayashi2004programmable}, a natural sensor that degrades $\lambda$CI through RecA, as in the $\lambda$-bacteriophage switch \cite{ackers1982quantitative}. In response to exposure to ultraviolet light or mitomycin C, RecA cleaves the $\lambda$CI repressor, increasing the degradation of $\lambda$CI \cite{kobayashi2004programmable, munsky2010identification, tian2006stochastic, wang2007noise}. Changes caused by a decrease in gene expression levels when exposed to DNA damage can create an environment for cells to transition between the two steady states of the system, shown both experimentally and theoretically \cite{kobayashi2004programmable, munsky2010identification, tian2006stochastic, wang2007noise}.}

%% mapping to real world system & study extensions
We have deliberately considered a highly simplified and generalisable model for cellular decision making, in an attempt to maximise the breadth of our findings. Several points must be considered in mapping these results to a specific biological system. First, the free energy for cellular reactions fuelled by ATP comes from the ATP:ADP ratio, which we here coarse-grain into a single effective energy availability parameter $A^*$. Next, this study concentrated on a toggle-switch type architecture consisting of two genes with feedback loops. \sj{As discussed above, this} motif consisting of two genes with feedback loops is found at the core of several known decision-making architectures throughout biology, but of course, it will not exist in isolation in real biological systems (although theoretical studies have underlined how analysis of its dynamics can be valuable in understanding its downstream physiological significance \cite{huang2007bifurcation, johnston2018identification, johnston2012mitochondrial}). Natural extensions to this work would include considering additional species within the gene architecture, which will expand the space of states accessible to the cell and begin to represent the large number of possible phenotypes supported by specific regulatory modules in biology. The influence of noise in gene expression \cite{paulsson2004summing, pedraza2005noise, thattai2001intrinsic} on the behaviour and stability of this system will also be valuable to link to specific biological situations. Together, the inclusion of noise and additional genetic actors will help shed further light on how decisions are made by the cell (in our model, how transitions between available attractor states are accomplished). Some subtle and re-entrant behaviour was observed in the bifurcation dynamics associated with transitions between different numbers of distinct stable states. If this behaviour is general across other architectures, it may provide a potential optimum intracellular energy range for phenotypic diversity, enabling a cell to have superior adaption to stochastic extracellular environmental change. 

%% closing statement
We hope that this study has opened a new line of enquiry in the current understanding of cellular decision making. Prior to this work, energy variability in cells had, to our knowledge, not been considered as a key determinant of the cellular decision-making landscape. Through consideration of the potential effect of energy variability on intracellular physiology, understanding the mechanisms behind cellular decision-making may improve. We hope that further knowledge on how intracellular energy budget changes a cell's decision-making landscape could be a step towards understanding the fundamental mechanisms behind cellular decision-making and towards developing novel methods to either promote or inhibit the consequences.

\section*{Data availability}
All scripts used for this study are openly accessible through:
\url{https://github.com/StochasticBiology/energy-variability-decision-making}.

\section*{Acknowledgements}
\igj{RK acknowledges PhD funding from the Wellcome Trust. IGJ acknowledges ERC grant 805046 (EvoConBiO) and a Turing Fellowship from the Alan Turing Institute.}
\sj{SJ thanks the BBSRC for grant code BB/M021386/1.}

\bibliographystyle{unsrt}
\bibliography{refs}

\newpage
\section*{Supplementary information}
\beginsupplement

\subsection*{Modelling the Energy Parameter Through a Linear Function}
\begin{figure}[h]
	\centering
	\includegraphics[width=0.85\textwidth]{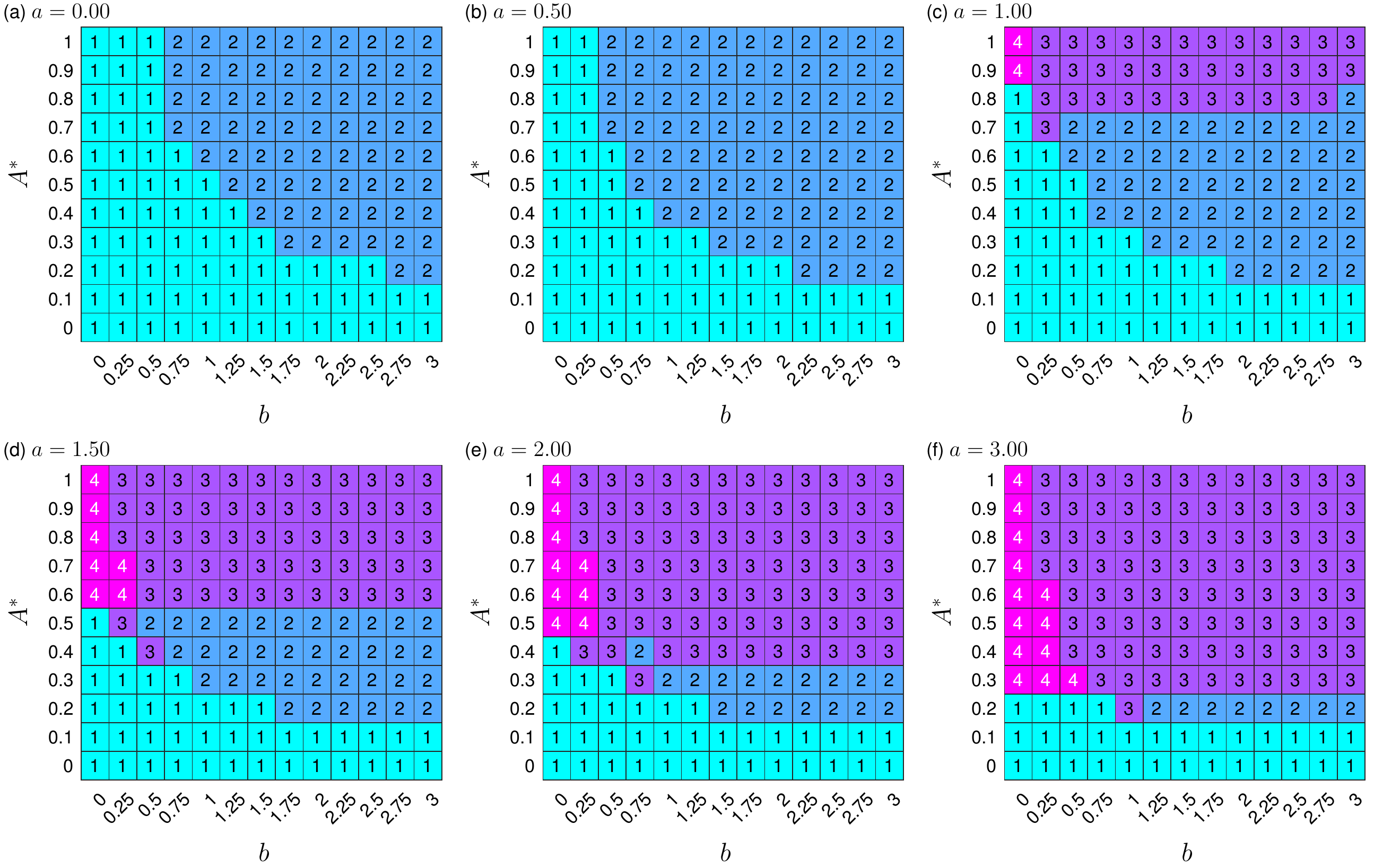}
	\caption{\textbf{Linear energy function displays similar qualitative decision-making landscapes to the default case.} Panels (a)-(f) display heatmaps for $6$ increasing values of $a$, when $\lambda = A^{*}$. Each panel exhibits the number of stable steady states for combinations of $b \in [0, 3]$ and $A^{*} \in [0, 1]$\, with all remaining parameters fixed at their default values.}
	\label{fig:hm-linear}
\end{figure}

The parameter $\lambda$ was initially modelled by the sigmoidal curve in Equation 4. Modelling the ATP-dependent modulatory term through a linear relationship, $\lambda = A^{*}$ (Fig. \ref{fig:hm-linear}), whilst fixing $\theta_a$, $\theta_b$, $n$ and $k$ at their default value, shows qualitatively similar behaviour compared to when $\lambda(A^{*})$ is sigmoidal (Fig. 4). In contrast we observe that significantly lower $A^{*}$ thresholds are required for multiple stable steady states, increased parameter sets with $4$ stable attractors and there exist more parameter sets with re-entrant behaviour.

The choice of function to model $\lambda$ therefore has an effect on the qualitative behaviour, but the number of stable attractors still increases as $A^{*}$ increases. Biologically this again displays an increased decision-making landscape for a cell as intracellular energy budget increases, as observed for sigmoidal $\lambda$.

\subsection*{Sigmoidal Energy Dependent Parameter}
\begin{figure}[h]
	\centering
	\includegraphics[scale=0.7]{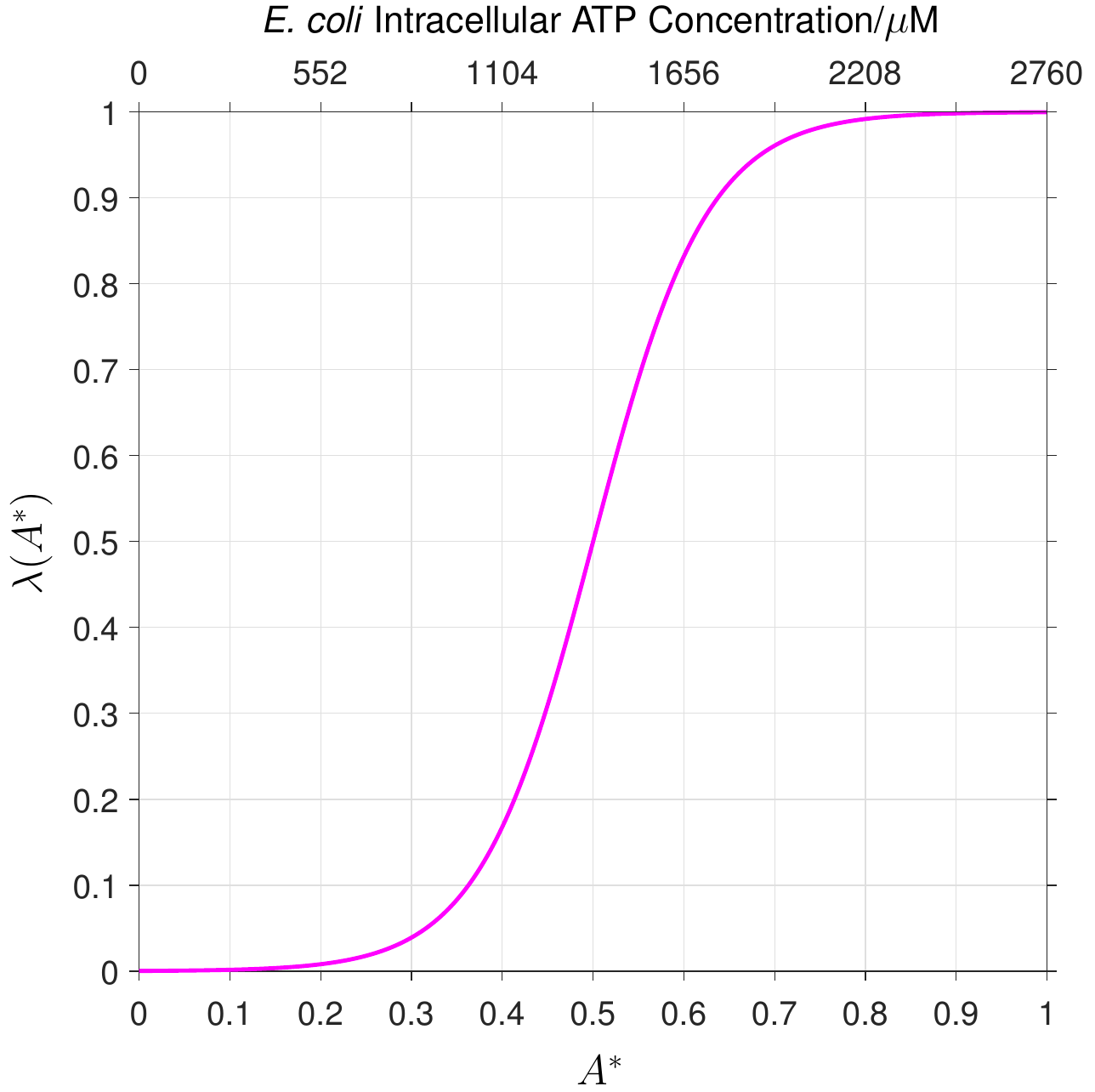}
	\caption{\textbf{Model relationship between energy availability and gene expression rates.} Sigmoidal curve given by Equation 4, displaying limiting $\lambda$ for small $A^{*}$ and maximal $\lambda$ as $A^{*}$ approaches $1$. Secondary $x$-axis (top $x$-axis) provides a biological example of the intracellular ATP concentration of \textit{E. coli} with respect to the maximum ATP concentration using known upper and lower bounds \cite{yaginuma2014diversity}. In this example, the lower bound of \textit{E. coli} intracellular ATP concentration is $320~\mu$M, corresponding to small $\lambda$.}
	\label{fig:sigmoid}
\end{figure}

The energy dependent parameter $\lambda(A^{*})$, modelled through Equation 4 with  $s_1=16$ and $s_2=-8$, is displayed in Fig. \ref{fig:sigmoid}. Included is a biological example, displaying the known \textit{E. coli} intracellular ATP concentration range ($1.54 \pm 1.22$~mM \cite{yaginuma2014diversity}).

\newpage
\subsection*{Bifurcation Diagram Example}
\begin{figure}[h]
	\centering
	\includegraphics[width=\textwidth]{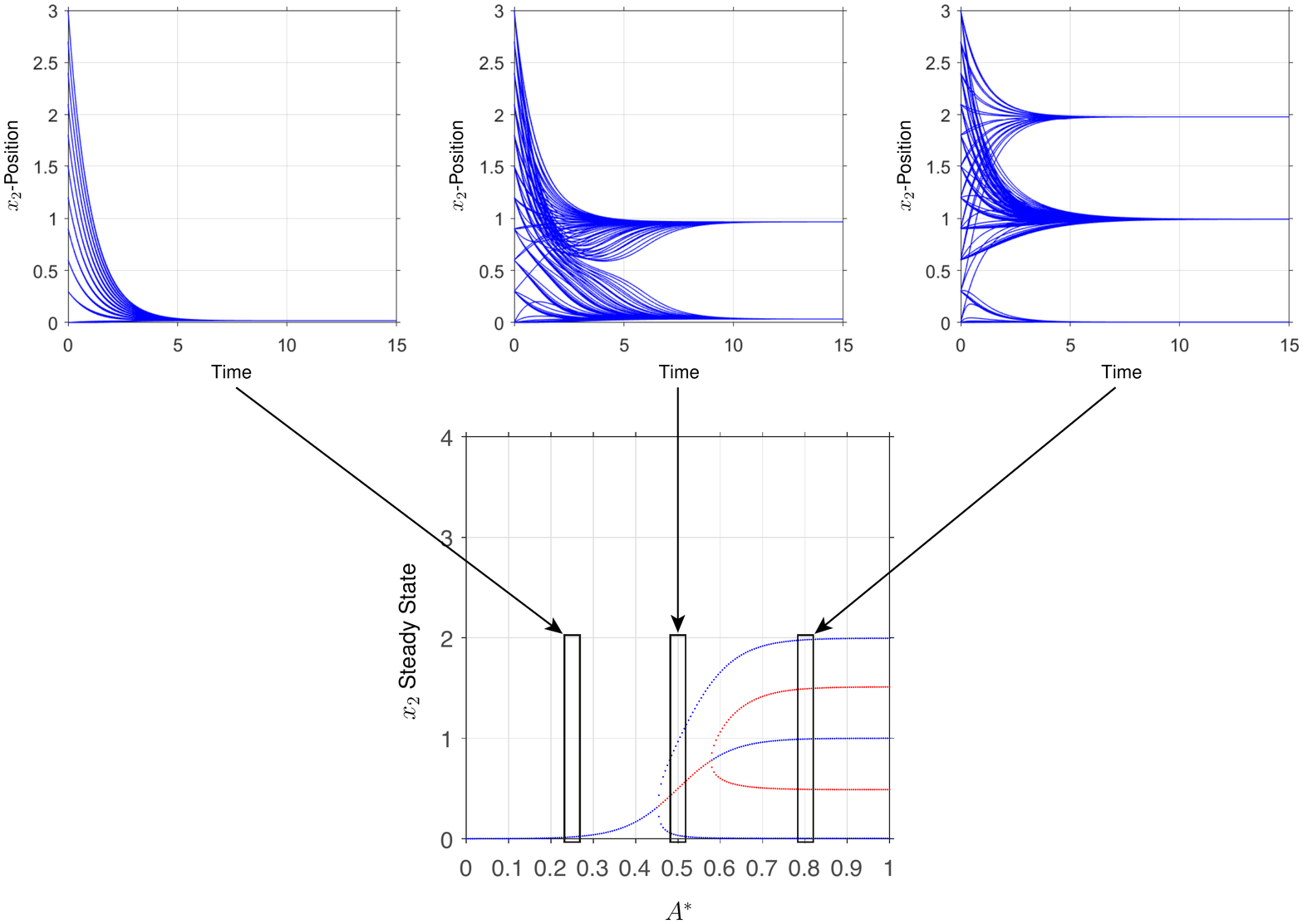}%could have width=x\linewidth or x\textwidth
	\caption{\textbf{Increasing energy levels support more distinct attractors.} Bifurcation diagram for the default parameter set, displaying stable steady states (blue circles) and unstable steady states (red circles) for $A^{*} \in [0, 1]$ in steps of size $5\times 10^{-3}$. Insets show behaviour of time-dependent simulations with multiple initial conditions at specified $A^{*}$ values. Time-dependent examples are displayed for $A^{*} = 0.25, 0.5$ and $0.8$ to show examples of the behaviour included in the bifurcation diagram.}
	\label{fig:bif_examples}
\end{figure}

Figure \ref{fig:bif_examples} displays the bifurcation diagram for default values and varying $A^*$. Inset figures show time-dependent trajectories towards stable attractors for fixed $A^{*}$ values.

%\textcolor{red}{An example bifurcation diagram with the default parameter values is shown in Figure \ref{fig:bif_examples}. Inserted sub-figures display time evolution of the $x_2$-position for each fixed $A^{*}$ until stable steady state is reached over a range of initial conditions and the type of steady state behaviour contained within each bifurcation diagram. The time-dependent simulations for each fixed $A^{*}$ value were simulated for equivalent grids of initial conditions, spanning $x_1$ and $x_2$ in equal steps.}

\newpage
\subsection*{Re-Entrant Behaviour}
\begin{figure}[h]
	\centering
	\includegraphics[width=.4\textwidth]{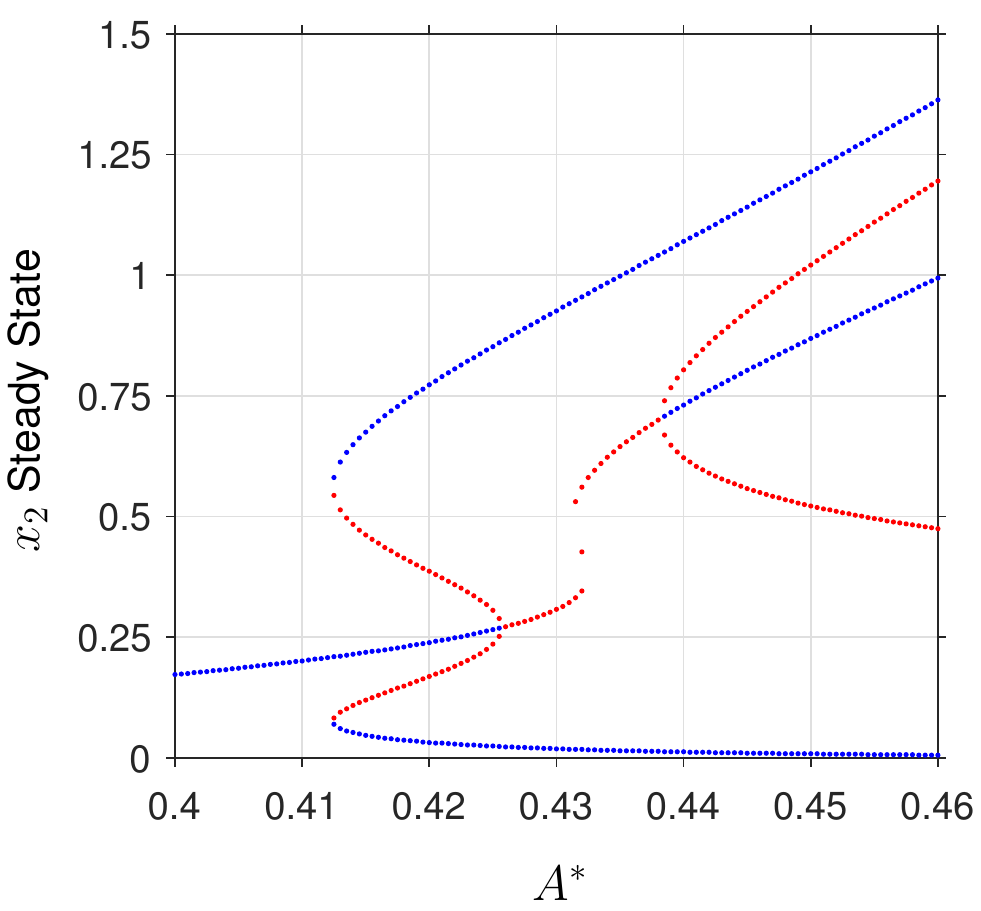}
	\caption{\textbf{Re-entrant behaviour in decision making options.} Close-up view of Figure 2(d) over $A^{*} \in [0.4, 0.46]$ in steps of size $5\times 10^{-4}$. The figure displays stable steady states (blue circles) and unstable steady states (red circles). The number of stable steady states transitions from $1$ to $3$, then $3$ to $2$ and finally $2$ to $3$ in a small region of $A^{*}$.}
	\label{fig:bif_a-zoom}
\end{figure}

\begin{figure}[htbp]
	\centering
	\includegraphics[width=.4\textwidth]{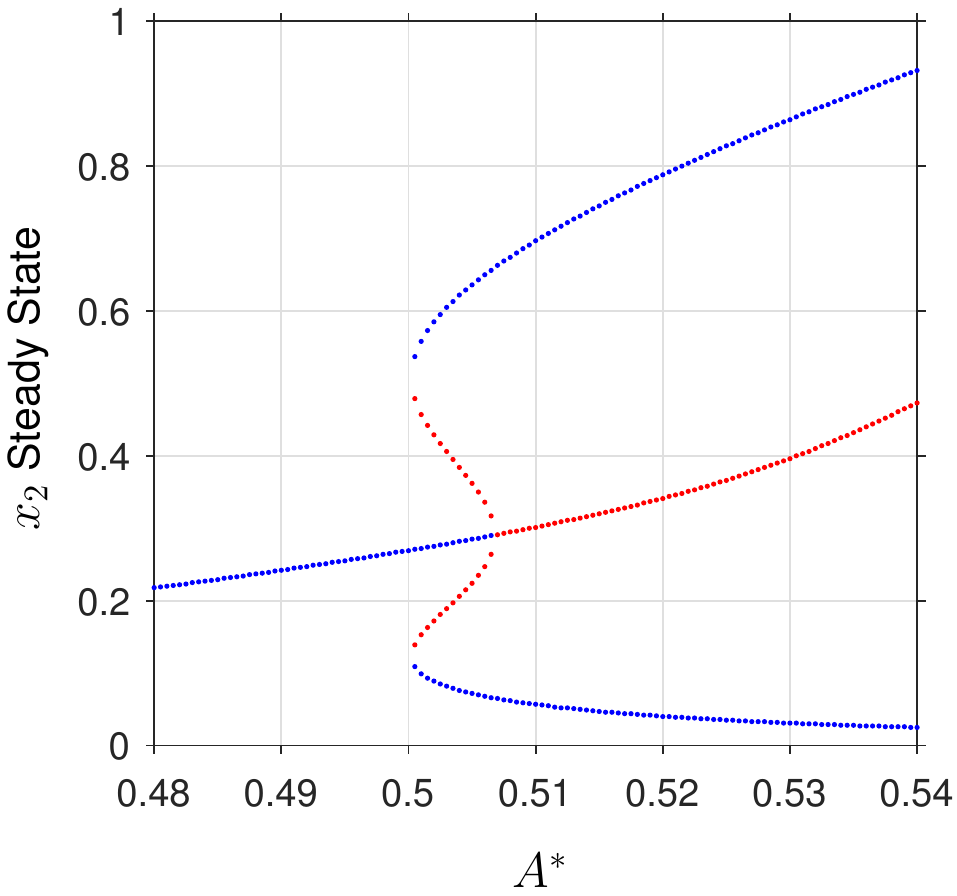}
	\caption{\textbf{Destabilisation of central attractor in re-entrant behaviour.} Close-up view of Figure 2(f) for $A^{*} \in [0.45, 0.5]$ in steps of size $5\times 10^{-4}$. The figure displays stable steady states (blue circles) and unstable steady states (red circles). The number of stable steady states transitions from $1$ to $3$, then $3$ to $2$ in the small region of $A^{*} \in [0.477, 0.484]$.}
	\label{fig:bif_b-zoom}
\end{figure}

\begin{figure}[htbp]
	\centering
	\includegraphics[scale=1]{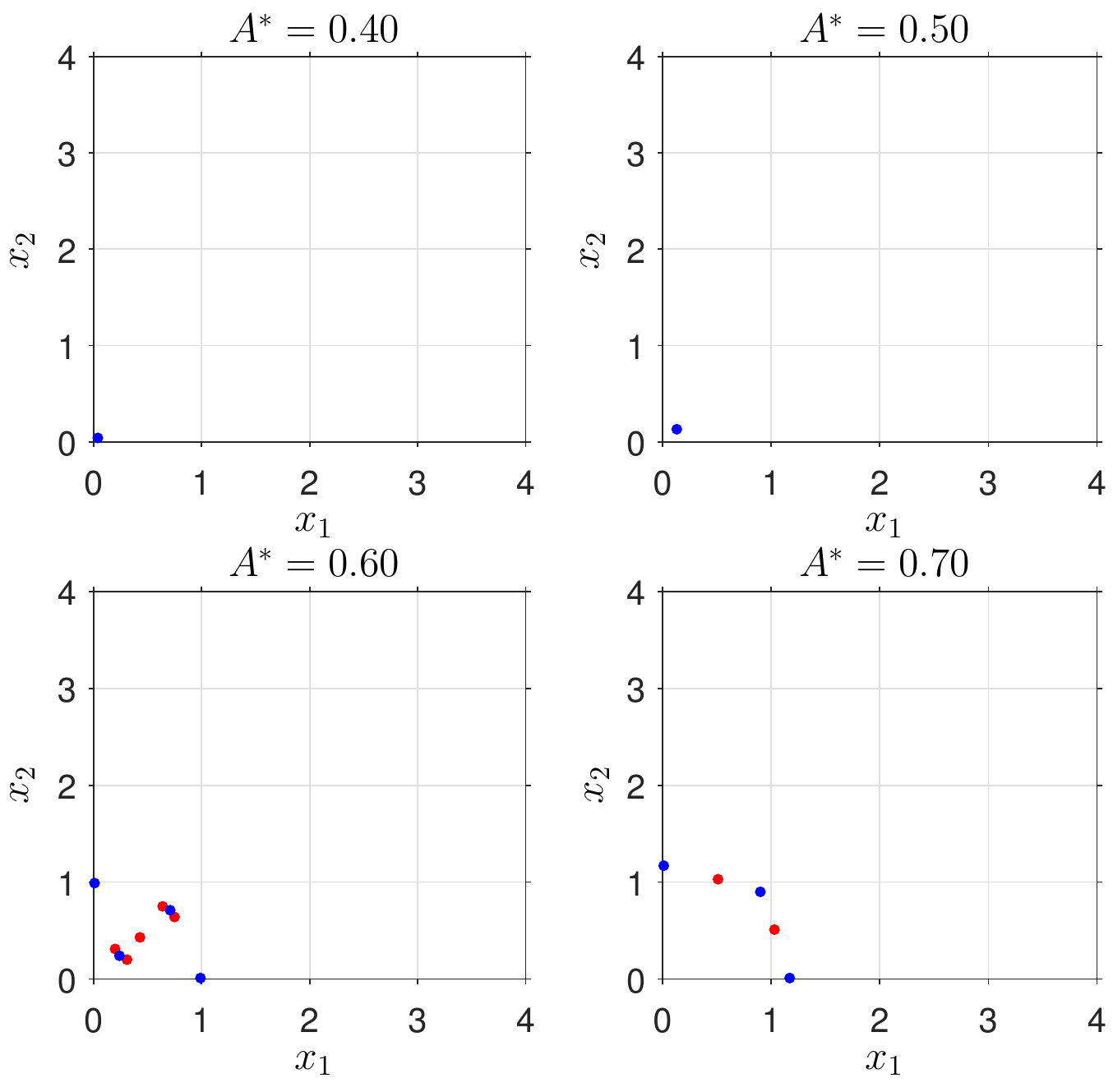}
	\caption{\textbf{Re-entrant behaviour showing destabilisation of the `near-zero' attractor.} Array displays stable (blue) and unstable (red) states for $a = 1$, $b = 0.25$ and $A^{*} \in [0.4, 0.7]$ in Figure 4(c). Increasing energy develops $4$ stable attractors which decreases to $3$ as $A^{*}$ increases. All remaining parameters are at their default value.}
	\label{fig:array-1}
\end{figure}

\begin{figure}[htbp]
	\centering
	\includegraphics[scale=1]{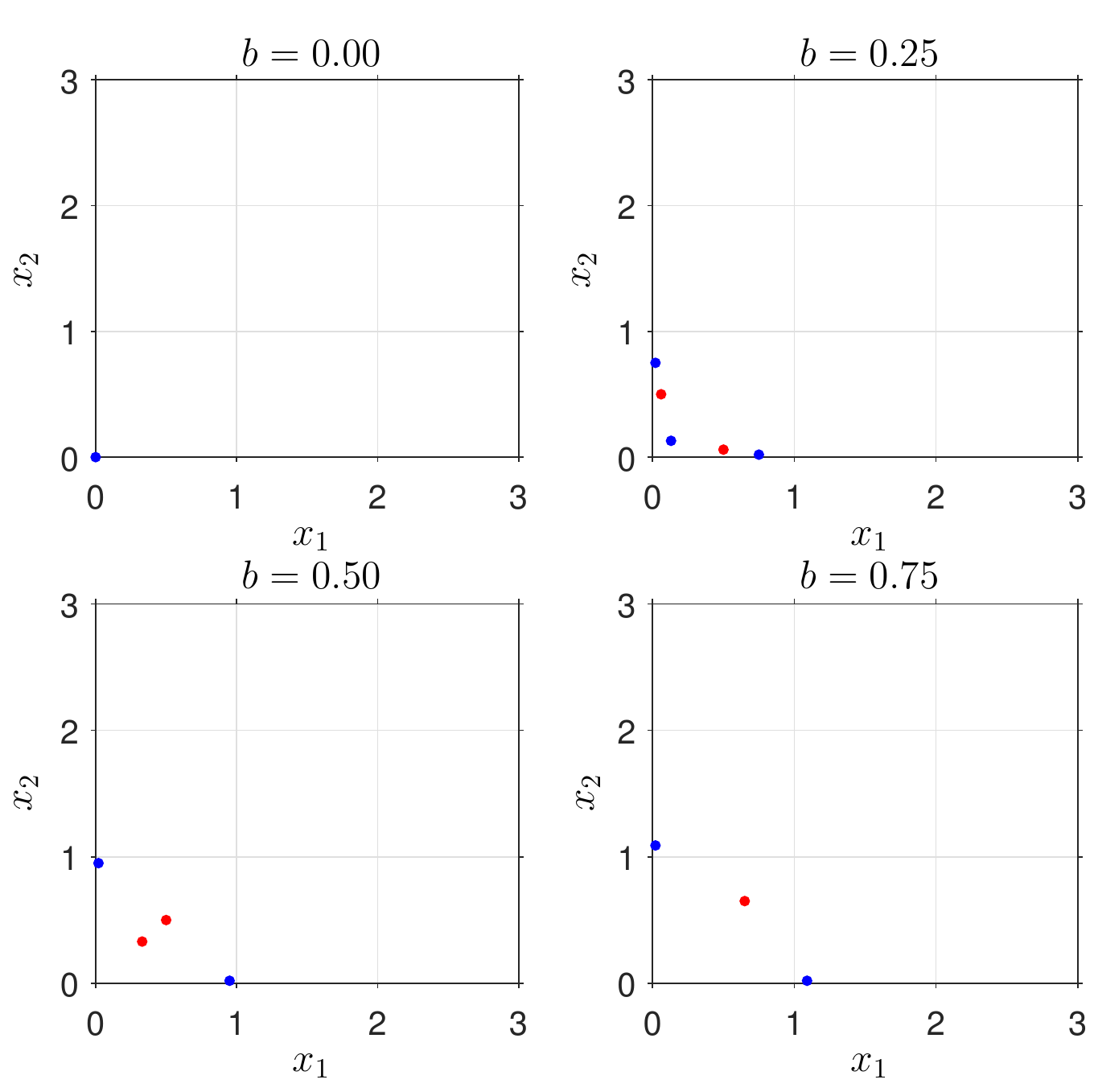}
	\caption{\textbf{Re-entrant behaviour showing destabilisation of the central attractor.} Array displays stable (blue) and unstable (red) states for $a = 1.5$, $A^{*} = 0.5$ and $b \in [0, 0.75]$ in Figure 4(d). Increasing energy develops $3$ stable attractors from the initial single stable state which decreases to $2$ as $b$ increases. All remaining parameters are at their default value.}
	\label{fig:array-2}
\end{figure}

Figures \ref{fig:bif_a-zoom} and \ref{fig:bif_b-zoom} show the re-entrant behaviour displayed in Figs. 2(d) and (f), respectively. Re-entrant behaviour contained in Figs. 4(c)-(d) are presented in Fig. \ref{fig:array-1} (varying $A^{*}$) and \ref{fig:array-2} (varying $b$), respectively.

\newpage
\subsection*{Effects on the Decision-Making Landscape When Reducing $n$}
\begin{figure}[htbp]
	\centering
	\includegraphics[width=\textwidth]{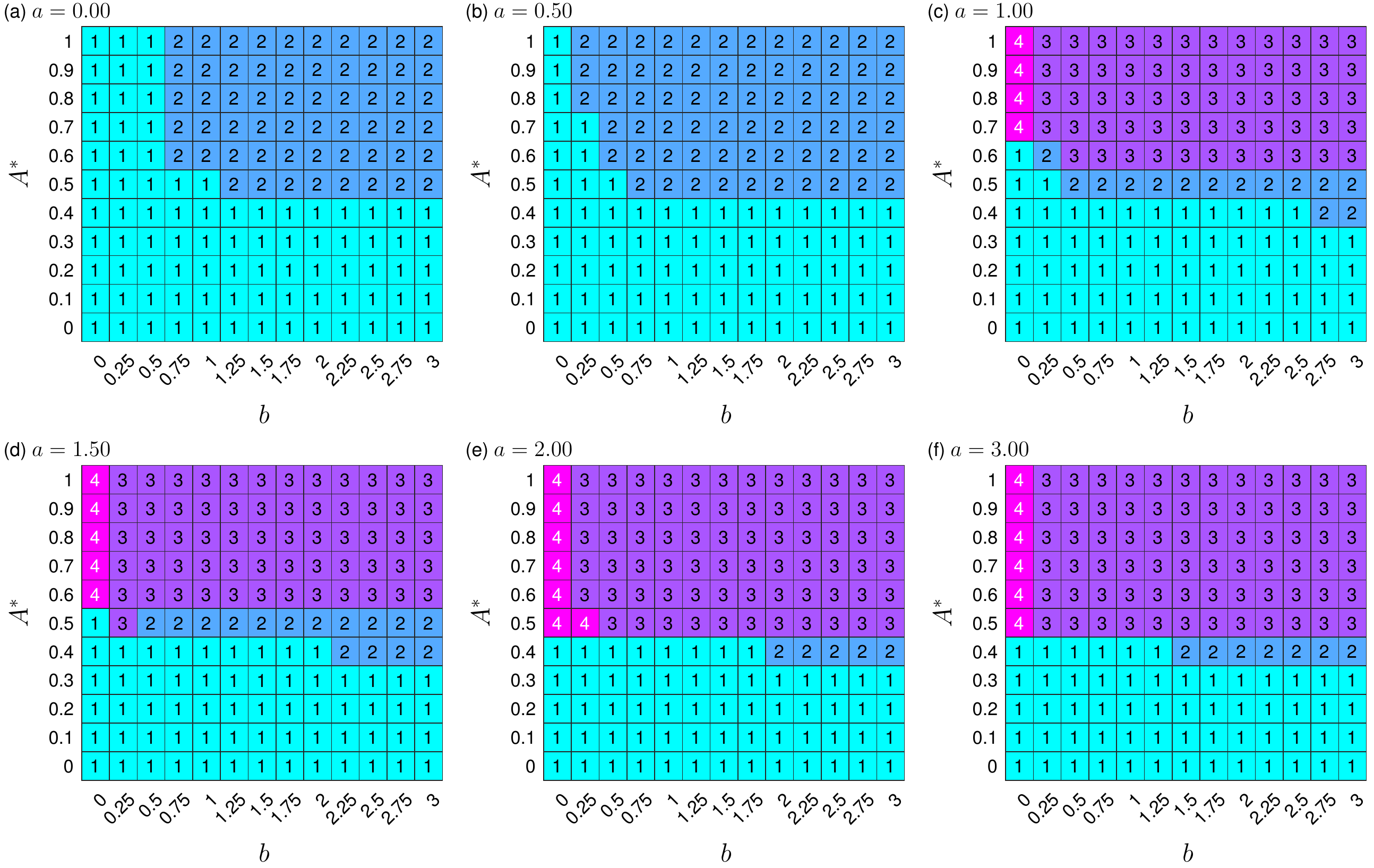}
	\caption{\textbf{Reduced Hill coefficient value ($n = 3$) displays similar qualitative decision-making landscapes to the default value ($n = 4$).} Panels (a)-(f) display heatmaps for $6$ increasing values of $a$, when $n=3$. Each panel exhibits the number of stable steady states for combinations of $b \in [0, 3]$ and $A^{*} \in [0, 1]$, with all remaining parameters fixed at their default values.}
	\label{fig:hm-n3}
\end{figure}

\begin{figure}[htbp]
	\centering
	\includegraphics[width=\textwidth]{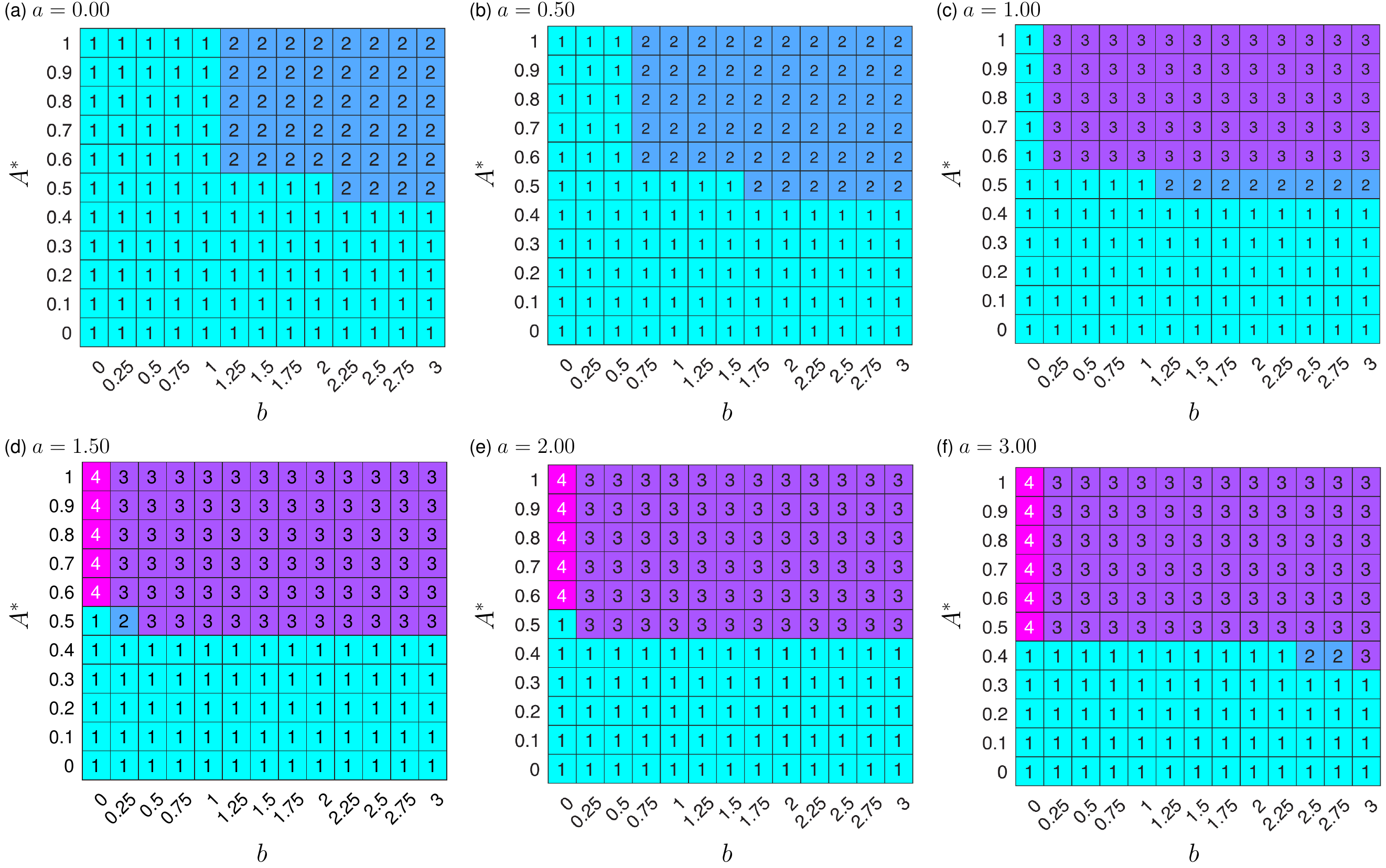}
	\caption{\textbf{Hill coefficient $n = 2$ displays simplified qualitative decision-making landscapes to the default value ($n = 4$).} Panels (a)-(f) display heatmaps for $6$ increasing values of $a$, when $n=2$. Each panel exhibits the number of stable steady states for combinations of $b \in [0, 3]$ and $A^{*} \in [0, 1]$, with all remaining parameters fixed at their default values.}
	\label{fig:hm-n2}
\end{figure}

\begin{figure}[htbp]
	\centering
	\includegraphics[width=\textwidth]{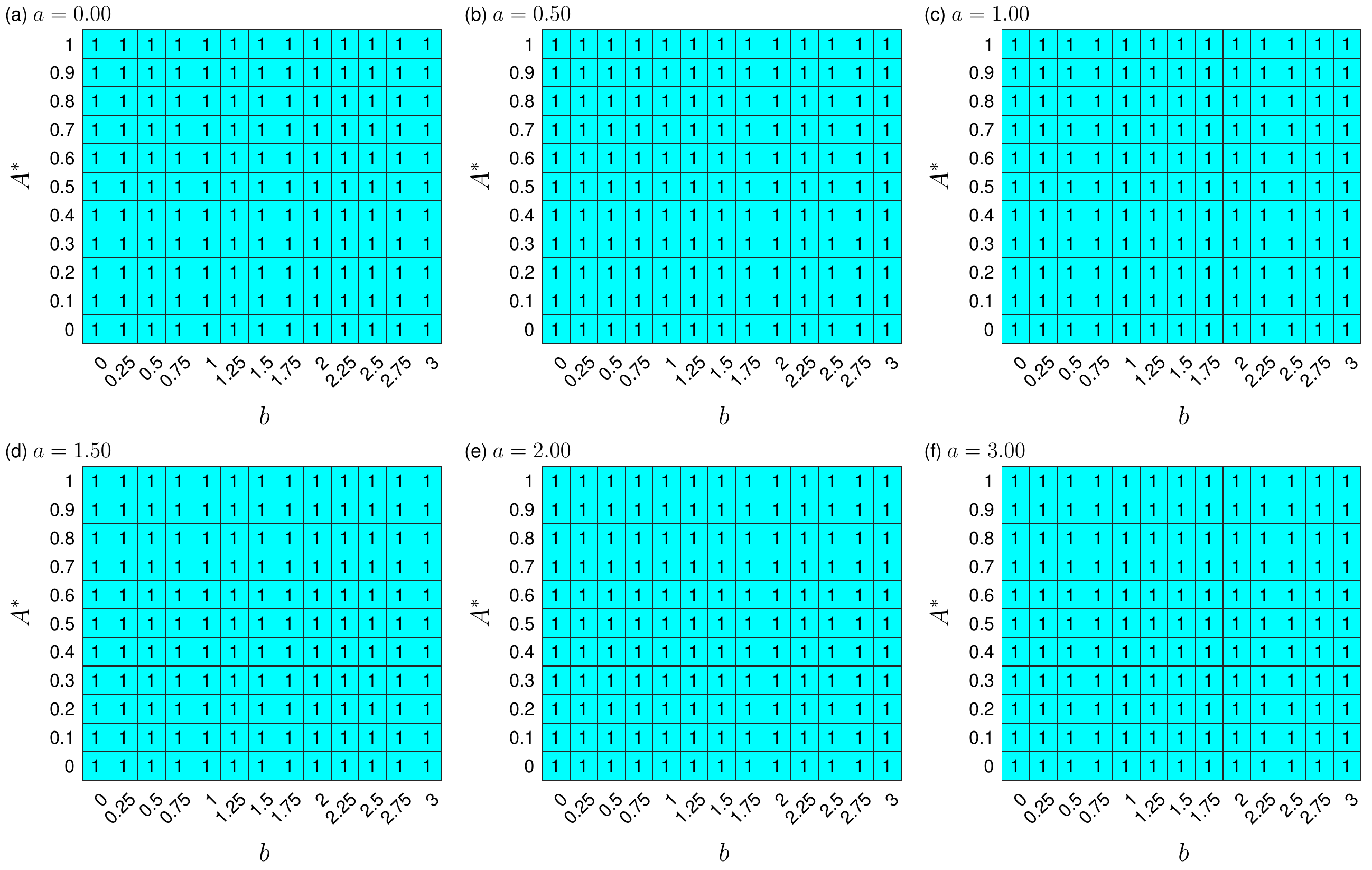}
	\caption{\textbf{Non-cooperative binding limits the decision-making landscape.} Panels (a)-(f) display heatmaps for $6$ increasing values of $a$, when $n=1$. Each panel displays the number of stable steady states when spanning $b \in [0, 3]$ and $A^{*} \in [0, 1]$, with all remaining parameters fixed at their default values.}
	\label{fig:hm-n1}
\end{figure}

Decreasing the Hill function coefficient, $n$, from the default value ($n=4$) to $n=1$ is displayed in Figs. \ref{fig:hm-n3}-\ref{fig:hm-n1} and exhibit the changes in qualitative behaviour as the amount of cooperativity is reduced.

\newpage
\subsection*{Regulatory Protein Binding Strength Heatmaps}
\begin{figure}[htbp]
	\centering
	\includegraphics[scale=0.8]{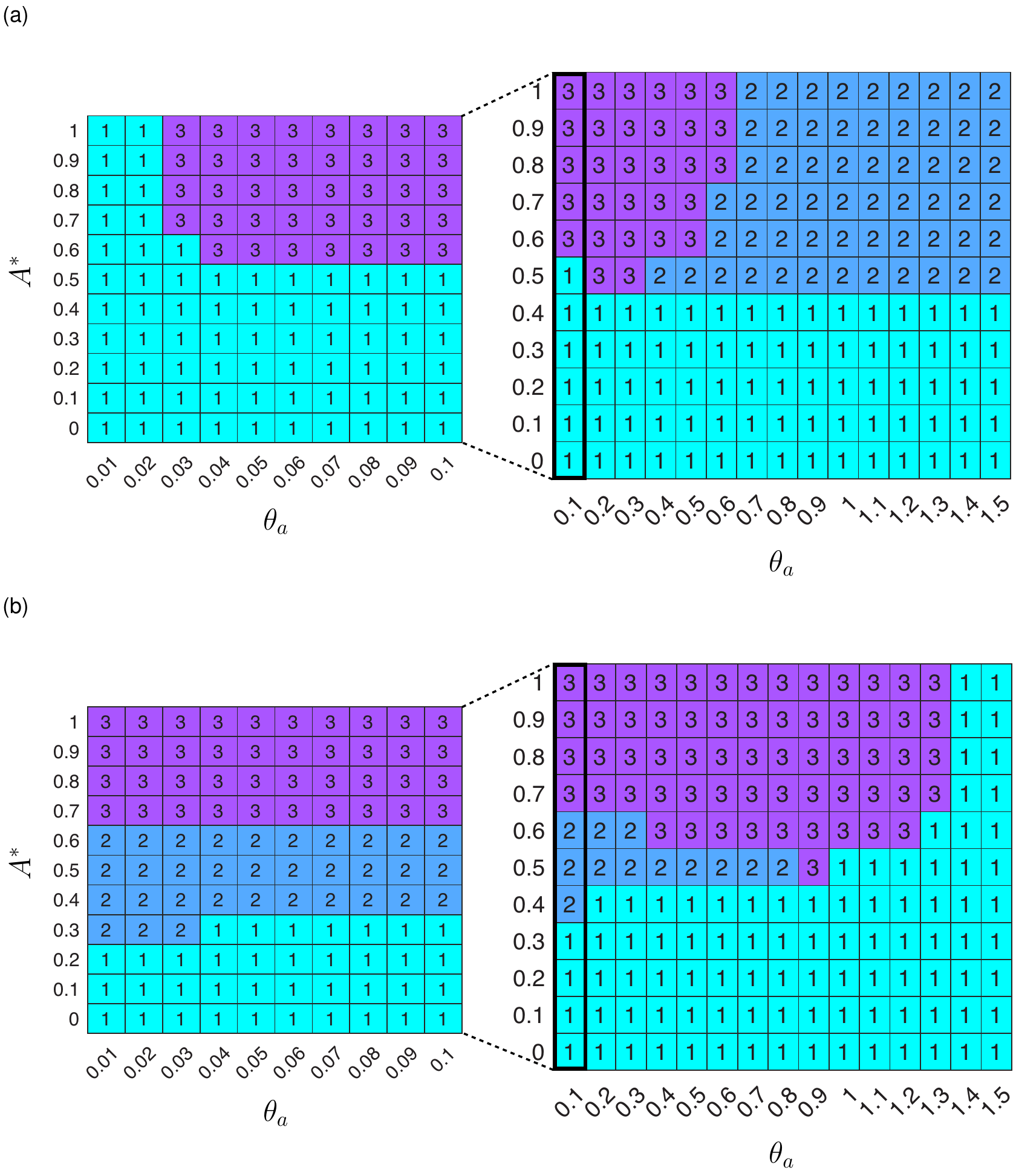}
	\caption{\textbf{Strength of binding limits the decision-making landscape.} Main panels display the number of stable states for (a) $\theta_a$, (b) $\theta_b \in [0.1, 1.5]$ and $A^{*} \in [0, 1]$. Inset sub-panels display stable attractor behaviour for $\theta_a$, $\theta_b \in [0.01, 0.1]$. Each panel contains colour coordinated number of stable steady states: $1$ (turquoise); $2$ (blue); $3$ (purple). Remaining parameters are fixed at their default values. Behaviour displayed for $\theta_a = \theta_b = 1.5$ continues for larger values of each parameter.}
	\label{fig:theta-ab}
\end{figure}

Figures \ref{fig:theta-ab}(a)-(b) show the effect of varying activator and repressor protein binding strength, respectively.

%% SI end

\end{document}